\documentclass{emulateapj}

\def\wise{{\it WISE }}

\begin{document}

\shortauthors{Luhman \& Mamajek}
\shorttitle{Disk Population of Upper Scorpius}

\title{The Disk Population of the Upper Scorpius Association\altaffilmark{1}}

\author{
K. L. Luhman\altaffilmark{1,2} and
E. E. Mamajek\altaffilmark{3,4}
}

\altaffiltext{1}{Department of Astronomy and Astrophysics, The Pennsylvania
State University, University Park, PA 16802; kluhman@astro.psu.edu.}
\altaffiltext{2}{Center for Exoplanets and Habitable Worlds,
The Pennsylvania State University, University Park, PA 16802.}
\altaffiltext{3}{Cerro Tololo Inter-American Observatory,
Casilla 603, La Serena, Chile}
\altaffiltext{4}{On leave, Department of Physics and Astronomy,
The University of Rochester, Rochester, NY 14627.}

\begin{abstract}

We present photometry at 3--24~\micron\ for all known members of the Upper
Scorpius association ($\tau\sim11$~Myr) based on all images of these objects
obtained with the {\it Spitzer Space Telescope} and the {\it Wide-field
Infrared Survey Explorer}. We have used these data to identify the members that
exhibit excess emission from circumstellar disks and estimate the evolutionary
stages of these disks. Through this analysis, we have found $\sim50$
new candidates for transitional, evolved, and debris disks.
The fraction of members harboring inner primordial disks is
$\lesssim10$\% for B--G stars ($M>1.2$~$M_\odot$) and increases with later
types to a value of $\sim25$\% at $\gtrsim$M5 ($M\lesssim0.2$~$M_\odot$),
in agreement with the results of previous disk surveys of smaller 
samples of Upper Sco members. These data indicate that the lifetimes
of disks are longer at lower stellar masses, and that a significant fraction
of disks of low-mass stars survive for at least $\sim10$~Myr. Finally, we
demonstrate that the distribution of excess sizes in Upper Sco and
the much younger Taurus star-forming region ($\tau\sim1$~Myr) are
consistent with the same, brief timescale for clearing of inner disks.

\end{abstract}


\section{Introduction}
\label{sec:intro}

Identifying the stars that harbor circumstellar disks within young
clusters and associations is an essential step in using disks to
study star and planet formation.
For instance, thorough surveys for disks provide well-defined
samples of targets for detailed measurements of disk properties.
In addition, by comparing the prevalence of disks in different evolutionary
stages within a given cluster and among clusters across a range of ages,
one can estimate the lifetimes and clearing timescales of disks, which
in turn constrain models of disk evolution and planet formation.

The best available tool for disk surveys has been mid-infrared (IR)
imaging with the {\it Spitzer Space Telescope} \citep{wer04}.
{\it Spitzer} has offered the sensitivity to detect the
photospheres of young brown dwarfs in nearby clusters ($d<500$~pc),
the field of view to efficiently map these clusters ($\sim5\arcmin$),
the angular resolution to resolve most of their members ($\sim2\arcsec$),
and a suite of mid-IR filters that could roughly classify
the evolutionary stages of disks (3.6--24~\micron).
Although {\it Spitzer} has focused on relatively compact clusters
\citep[$\sim1$~deg$^2$,][]{lada06,sic06,dah07,her07,luh08cha,gut09},
it has also imaged more widely distributed populations in areas as large as
$\sim50$~deg$^2$ \citep{eva09,reb10}.

The Upper Scorpius subgroup in the Scorpius-Centaurus OB association
is an attractive target for disk surveys.
It contains one of the nearest and richest populations of young stars
\citep[$d=125$--165~pc, $N\sim800$,][]{pm08}, making it possible to measure
disk properties down to low stellar masses with good statistical accuracy.
Perhaps most importantly, Upper Sco is at an age where both primordial
disks and second-generation debris disks are abundant \citep{car09}, and thus
can potentially provide vital constraints on disk evolution.
Most studies have adopted a mean age of 5~Myr for Upper Sco
\citep{deg89,pre02}, but \citet{pec12} recently
derived a significantly older value of 11~Myr, which affects 
the timescales of disk evolution derived from any census of disks in
the region. The ages of Upper Sco and its neighboring subgroups remain a
subject of debate \citep{son12}.

{\it Spitzer} has been used to identify and classify disks
in Upper Sco based on mid-IR photometry \citep{car06,car09,chen05,chen11,ria09}
and to study the mineralogy of dust in those disks with mid-IR spectroscopy
\citep{sch07,dah09}.
In one of the most widely-cited results from the {\it Spitzer} data,
\citet{car06} found that, in a sample of $\sim200$ members, nearly all of the
stars more massive than the Sun have lost their primordial disks while a
significant fraction of the low-mass stars ($\sim$1/5) have retained them,
indicating that disk lifetimes may be longer at lower stellar masses. 
However, because Upper Sco is spread across a large area ($\sim100$~deg$^2$),
{\it Spitzer} imaged members individually rather than through a map of
the association, and only a subset of the stellar population was observed.
As a result, the potential of the association for constraining the
properties of disks was not fully exploited.

The {\it Wide-field Infrared Survey Explorer} \citep[{\it WISE},][]{wri10}
has presented an opportunity to significantly improve the census of disks
in Upper Sco. In 2010, the satellite imaged the entire sky at 3.4, 4.6, 12,
and 22~\micron, providing mid-IR photometry for the areas of Upper
Sco that were not observed by {\it Spitzer}.
Although they have lower sensitivity and angular resolution than {\it Spitzer}
images, the \wise data do detect every known member of Upper Sco (that
is resolved) in the most sensitive bands, including the brown dwarfs.
\wise photometry has already been used to search for disks around members of
Upper Sco at B, A, and F types \citep{riz12} and mid- to late-M types
\citep{ria12}.

To obtain the most thorough census of disks in Upper Sco that is possible
with available data, we have compiled all photometry between
3--24~\micron\ from both {\it Spitzer} and \wise for all known members of the
association (Section~\ref{sec:data}).
We use these data to identify the members that exhibit
significant mid-IR excesses (Section~\ref{sec:exc}) and to estimate the
evolutionary stages of the detected disks (Section~\ref{sec:class}).
We then measure the distribution of excesses as a function of spectral type,
wavelength, disk class, and excess size and discuss the resulting
implications for disk evolution (Section~\ref{sec:global}).

\section{Data for Upper Sco}
\label{sec:data}

\subsection{Known Members}
\label{sec:mem}

To characterize the population of disks in Upper Sco, we began by
constructing a list of all published sources that are likely to be members
of the association based on proper motions or diagnostics of youth.
We also have included new members that we have uncovered through an ongoing
survey of the association. The basis for inclusion in our membership
catalog will be described in a separate study (K. Luhman, in preparation).
We have treated known binaries as single sources if they are unresolved by both
the Two Micron All-Sky Survey \citep[2MASS,][]{skr06} and the United Kingdom
Infrared Telescope (UKIRT) Infrared Deep Sky Survey
\citep[UKIDSS,][]{law07}\footnote{UKIDSS uses the UKIRT Wide Field
Camera \citep[WFCAM,][]{cas07} and a
photometric system described by \citet{hew06}. The pipeline processing and
science archive are described by Irwin et al. (in preparation) and
\citet{ham08}.}.
Thus, secondaries that have been resolved only by high-resolution
imaging with the {\it Hubble Space Telescope} or adaptive optics 
do not receive separate entries in our catalog of members
\citep{luh05usco,kra05,laf08,laf11,bil11,ire11}.
Our resulting list of known members of Upper Sco contains 863 objects.
We present their names and spectral types in Table~\ref{tab:all}.
Our compilation of spectral classifications is not
fully exhaustive; additional measurements have been reported in the literature
for some of the stars, particularly at earlier types, but they generally
agree with the types in Table~\ref{tab:all}.

\subsection{Spitzer Photometry}
\label{sec:spitzer}

For our census of disks in Upper Sco, we make use of
images at 3.6, 4.5, 5.8, and 8.0~\micron\ obtained with
{\it Spitzer}'s Infrared Array Camera \citep[IRAC;][]{faz04} and
images at 24~\micron\ obtained with the Multiband Imaging Photometer for
{\it Spitzer} \citep[MIPS;][]{rie04}.
These bands are denoted as [3.6], [4.5], [5.8], [8.0], and [24].
The fields of view were $5\farcm2\times5\farcm2$ and
$5\farcm4\times5\farcm4$ for IRAC and the 24~\micron\ band of MIPS,
respectively. The cameras produced images with FWHM$=1\farcs6$--$1\farcs9$
from 3.6 to 8.0~\micron\ and FWHM=$5\farcs9$ at 24~\micron.

Several previous studies have reported {\it Spitzer} photometry for samples of
members of Upper Sco \citep{rie05,chen05,pad06,car06,car08,car09,sch07,ria09}.
In this work, we consider all IRAC and MIPS 24~\micron\ observations of 
all members in our catalog.
These data were obtained through Guaranteed Time
programs 19 (J. Houck), 40, 58 (G. Rieke), 72 (F. Low), and 84 (M. Jura),
Director's Discretionary Time program 248 (S. Mohanty),
Legacy programs 139, 173, 177 (N. Evans), 148 (M. Meyer), and 30574 (L. Allen),
and General Observer programs 20069 (J. Carpenter),
20103 (L. Hillenbrand), 20146 (K. Gordon), 20435 (R. Jayawardhana),
20803 (S. Mohanty), 30765 (K. Stapelfeldt), 50025 (P. Harvey),
50554 (I. Song), and 80187 (K. Luhman).
We have measured photometry from these images for the members of Upper Sco
with the methods described by \citet{luh10tau}.

We present our measurements of {\it Spitzer} photometry for Upper Sco in
Table~\ref{tab:all}. Some members were observed at multiple
epochs, most of which were separated by a few days. Since none of these objects
exhibit significant variability in these data, we report the mean of
multiple measurements weighted by the inverse square of their flux errors.
For each of the IRAC bands, the camera detected every member that appeared
within the field of view and was resolved.
This was not the case for the 24~\micron\ images
from MIPS. All of the non-detections in this band are indicated in
Table~\ref{tab:all}. The typical detection limit is
$[24]\sim10$.  We have not measured 24~\micron\ photometry for HIP~80425 and
HIP~80338 since they are dominated by extended emission.
2MASS J16105240-1937344 also may be significantly contaminated by extended
emission at 24~\micron. We report a tentative measurement for this star, but
it is excluded from the analysis in this paper.
\citet{car09} noted that their 24~\micron\ data for several additional stars
were unreliable because of extended emission.
Based on visual inspection of the images, we find that these stars are
point sources that are sufficiently bright compared to spatial variations in
the background emission so that our aperture photometry should be reasonably
accurate. This conclusion is supported by the fact that our 24~\micron\ data 
for these stars agree with the fluxes expected from the photospheres.
IRAC and MIPS observed 381 and 442 members of Upper Sco, respectively.
A total of 484 members were observed by at least one of the cameras,
corresponding to 56\% of the known members.
IRAC and MIPS data have not been previously published for 117 and 124 of
these sources, respectively, which includes 48 non-detections by MIPS.

\subsection{\wise Photometry}
\label{sec:wise}

In addition to the pointed {\it Spitzer} observations, we utilize mid-IR
photometry from the all-sky imaging survey by {\it WISE}.
The bands of \wise are centered at 3.4, 4.6, 12, and 22~\micron, which
are denoted as W1 through W4 \citep{wri10}. The angular resolution of
the \wise images is $\sim6\arcsec$ in the first three bands and
$\sim12\arcsec$ in W4.

We have retrieved the coordinates and profile-fit photometry of all objects
in the \wise Preliminary Release Source Catalog and the \wise 
All-Sky Source Catalog that are within a distance of $3\arcsec$ from
members of Upper Sco. 
For W1 and W2, most of the data agree within 5\% between the two catalogs,
and the catalogs exhibit comparable agreement with the {\it Spitzer} data.
The width of the sequence of diskless stars in $K_s-W1$
and $K_s-W2$ as a function of spectral type is also similar for
the two catalogs (Section~\ref{sec:exc}). We adopt
the W1 and W2 data from the All-Sky Source Catalog since it was
generated from a newer processing of the \wise images.
However, we find that the sequences of diskless stars in $K_s-W3$ and $K_s-W4$
are noticeably tighter with the data from the Preliminary Release Source
Catalog for the fainter stars, which implies better photometric
accuracies in this catalog for W3 and W4. 
This result is supported by a comparison of W4 to our measurements of [24];
the standard deviation of $W4-[24]$ is 0.18 and 0.28~mag using the
Preliminary Release and All-Sky data, respectively.
Therefore, we adopt the W3 and W4
measurements from the Preliminary Release Source Catalog, with the
exception of 2MASS J16045716-2104160, 2MASS J16133647-2327353,
and 2MASS J16133688-2327298, for which
measurements are available from only the All-Sky Source Catalog.
On average, the data from the Preliminary Release Source Catalog for Upper Sco
are 0.03~mag brighter in W3 and 0.035~mag fainter in W4 than the data
from the All-Sky Source Catalog.

We have modified our catalog of \wise data in several ways.
Since the W2 fluxes for saturated stars are significantly
overestimated \citep[0.2--1~mag at W2$<6$,][]{cut12},
we have omitted these measurements from our tabulation. These empty entries
in the catalog are labeled with the flag ``err".
Biases of this kind are also present in the other bands for the brightest
stars, but they are much smaller ($\lesssim0.2$~mag), so we retain these
measurements.
Although the All-Sky Source Catalog contains data for Antares, we label it
as saturated in our catalog since it is too bright for reliable photometry.
Through visual inspection of the \wise images, we have checked for members
of Upper Sco whose \wise photometry may be affected by the PSF of another
star. Each band in which this occurs is labeled with the flag ``bl"
(for ``blend"). We also have inspected the images to check
for unreliable detections and measurements that may be contaminated
by extended emission. For the former, we have omitted the
\wise photometry and have added the flag ``false" to our catalog.
The \wise photometry is labeled with the flag ``ext" if extended emission
is suspected based on the visual inspection.
If measurements are known to be dominated by extended emission based on
higher-resolution images from {\it Spitzer}, they are omitted from the catalog.
Some studies have excluded data that have high values of the reduced
$\chi^2$ for the profile fitting in the \wise catalog because this
may indicate the presence of extended emission \citep{riz12}.
However, we find that most of the members of Upper Sco with high $\chi^2$ do
not show evidence of extended emission in images from 
{\it Spitzer}, when available. Therefore, we rely on the steps that we have
described rather than the $\chi^2$ flag for identifying extended emission.
Finally, for the source in the \wise catalog matched to each member, we have
visually checked whether its position differs significantly ($\gtrsim2\arcsec$) 
among the four \wise bands. By doing so, we can identify erroneous matches in 
the \wise catalog between one object that dominates at shorter wavelengths and
a different object that dominates at longer wavelengths,
as in the case of the components of a binary system in which only the secondary
has a disk (e.g., 2MASS J16101888-2502325 and 2MASS J16101918-2502301) and 
stars and red galaxies with small projected separations
\citep[e.g., Sco PMS 17,][]{car09}.

We have examined the \wise images of all members that lack counterparts
in the \wise catalog.
Several of the unmatched members are not detected because of their proximity
to brighter stars. We label their empty entries with the flag ``unres".
Among the pairs separated by less than $6\arcsec$ (the FWHM for the
shorter bands), we have attached ``bin" (for ``binary") to the photometry
of the brighter components to indicate a possible contribution from the
fainter stars. 
As noted in Section~\ref{sec:mem}, many known close companions are not
included in our catalog of members. Thus, they represent additional unresolved
components in the \wise data that are not marked in our catalog.
Finally, through visual inspection of the images, we find that
\wise did detect four members that are absent from the Source Catalogs.
We found data for two of these stars, 2MASS J16071750-1820348
and 2MASS J16001730-2236504, in the \wise All-Sky Reject Table.
We measured photometry for the remaining stars, 2MASS J16090407-2417588 and 
2MASS J16081758-2348508, by applying aperture photometry to the \wise images. 
We present our catalog of \wise photometry for the known members of
Upper Sco in Table~\ref{tab:all}.

\section{Measurement of Infrared Excess Emission}
\label{sec:exc}

A dusty circumstellar disk within a few AU of a star is brightest
at IR wavelengths, becoming brighter than the star beyond $\sim5$~\micron.
As a result, one can detect the presence of a disk based on 
IR emission that is greater than that expected from a stellar photosphere.
We have applied this technique to the {\it Spitzer} and \wise photometry in
Upper Sco to identify the members that have disks.
To detect and measure excess emission for each member, we have
used the colors produced by the {\it Spitzer} and \wise bands relative to $K$
(2.2~\micron). We use $K$ for these colors because it is long enough in
wavelength so that extinction is low while short enough in wavelength
that it is usually dominated by the stellar photosphere in systems with disks.
For most stars, we adopt $K_s$ from the 2MASS Point Source Catalog or $K$ from
the seventh data release of UKIDSS, whose photometric systems are similar
\citep{hod09}. We use $K$ photometry from
\citet{the86} and \citet{kim04} for a few of the brightest members
since their 2MASS data have large uncertainties.
We focus on the {\it Spitzer} bands [4.5], [8.0], and [24] and
the \wise bands W2, W3, and W4 for detecting excess emission.
We do not consider [3.6] and W1 since
they should show the smallest excesses among the {\it Spitzer} and \wise bands.
The data at [5.8] is discussed only briefly since it is available for
a relatively small number of members.
Antares is excluded from our analysis since it is a red supergiant
and it lacks data from {\it Spitzer} or {\it WISE}.

The color excess of a star is measured by comparing its observed color
to the color expected for its stellar photosphere.
Because of its richness and moderately advanced age, Upper Sco contains
a large number of diskless stars, providing an accurate determination of
the colors of young photospheres.
Since photospheric colors can vary with spectral type, we plot the IR colors
in Upper Sco as a function of spectral type in Figure~\ref{fig:pan}.
The colors are not corrected for extinction, which is typically $A_V=1$--2
in this association. We have indicated in Figure~\ref{fig:pan}
the known and suspected Be stars \citep{mer33,jas64,cra68,cot93,cie10}
and the candidates for disks in different evolutionary classes that are
identified in Section~\ref{sec:class} and in previous studies
\citep{cie07,car09,chen06,chen11}.
For each color in Figure~\ref{fig:pan}, the data exhibit a narrow sequence of
blue sources and a broader distribution of redder sources, corresponding
to stellar photospheres and stars with disks, respectively.
We discuss the measurement of excesses from these colors in the
remainder of this section. The results of this analysis are summarized
in Table~\ref{tab:all}, where we indicate whether each member of Upper Sco
has excess emission in each of the six {\it Spitzer} and \wise bands
that we have considered.

\subsection{Excesses in [4.5] and W2}

We examine the data in [4.5] and W2 together
since these bands have similar effective wavelengths (4.5 and 4.6~\micron).
Photometry in [4.5] and W2 is available for 359 and 817 members of Upper Sco,
respectively. The stars that lack [4.5] were not observed by IRAC while
the stars that lack W2 are too bright for good photometry 
(Section~\ref{sec:wise}) or are blended with other sources.
Thus, all members were detected in both bands as long as they were
observed (by IRAC) and resolved.
26 members have [4.5] but not W2, which consist of stars that are
too bright for a reliable W2 measurement and companions that are resolved
by IRAC but not {\it WISE}. As a result, 843 members have data in at least one
of the two bands.

\begin{figure}
\epsscale{1}
\plotone{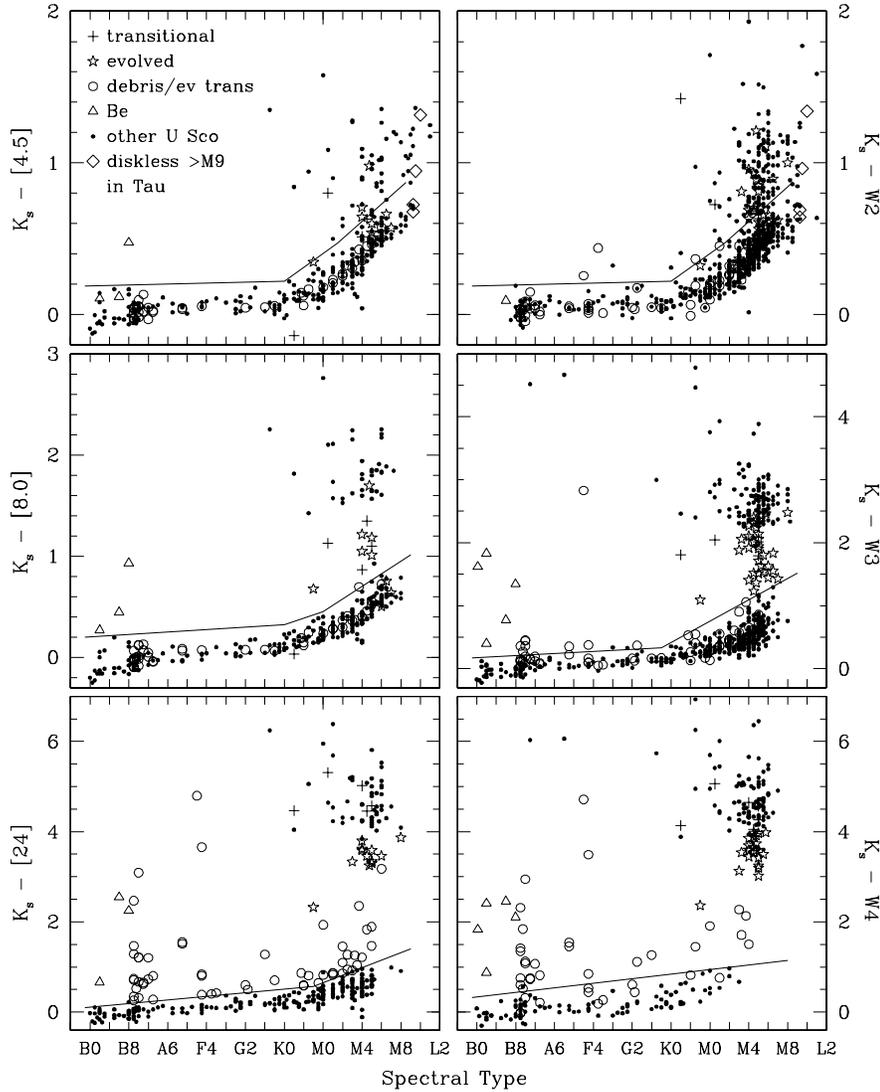}
\caption{
Infrared colors versus spectral type for members
of Upper Sco based on data from {\it Spitzer} (left) and \wise (right).
The solid lines are used to identify excess emission from circumstellar disks.
For types later than M8.5, 4.5~\micron\ and W2 excesses are difficult to
identify reliably because the photospheric colors increase rapidly,
as shown by the data for Upper Sco and late-type diskless members of
Taurus (diamonds).
We have indicated known Be stars \citep[triangles,][]{car09,cie10} and
candidates for transitional disks (crosses), evolved disks (stars), and debris
or evolved transitional disks \citep[circles,][this work]{car09,chen06,chen11}.
Because the photospheric
sequence is broader in $K_s-W4$ than $K_s-[24]$, some of the debris disk
candidates at 24~\micron\ appear below the excess boundary for W4. Data at
24~\micron\ and W4 that have errors larger than 0.25~mag or that appear
to be contaminated by extended emission are omitted.
}
\label{fig:pan}
\end{figure}

In the diagrams of $K_s-[4.5]$ and $K_s-W2$ versus spectral type
in Figure~\ref{fig:pan}, we have used the sequences of diskless stars to select
boundaries for identifying stars that exhibit significant color excesses. 
We use the same boundary for both colors since the data in [4.5] and W2
have an average offset of less than a few percent.
However, the photospheric sequences are not well-defined at the latest
spectral types because of the small number of members. 
The bluest L-type members have colors near 1.2, which is redder
than the photospheric values of $K_s-[4.5]$ estimated by \citet{luh10tau}
based on young L dwarfs in the solar neighborhood ($\tau\sim10$-100~Myr).
To better define the photospheric colors
for L-type members of Upper Sco, we include in Figure~\ref{fig:pan} 
data for members of Taurus that are later than M9 and that are diskless
based on photometry at longer wavelengths \citep{luh10tau}.
These data coincide with the bluest sources in Upper Sco and confirm the
rapid increase in $K_s-[4.5]$ and $K_s-W2$ beyond M9.
The boundaries for $K_s-[4.5]$ and $K_s-W2$ are defined by lines connecting
(B0, 0.19), (K0, 0.22), (M1.5, 0.47), and (M8.5, 0.87).
Since the dependence on spectral type of these colors remains somewhat
uncertain at M9--L2, we end the boundaries in Figure~\ref{fig:pan} at M8.5.
Measurements of excesses at the latest types are also sensitive to 
uncertainties in spectral types because of the rapid increase in the
photospheric colors.
Nevertheless, we have inspected the data to check for members later than M8.5
that show obvious excesses.
[LHJ2006] J163919.15-253409.9 has an excess in W2 but not [4.5].
This discrepancy is a reflection of the fact that W2 is brighter than
[4.5] by 0.35~mag. A similar offset is present between W1 and [3.6],
indicating that the source is variable. [LHJ2007] J160918.69-222923.7
exhibits the same characteristics. Mid-IR variability of this size
is more likely to occur in stars with disks \citep{luh08cha,luh10tau}.
Therefore, we classify these two objects as having excesses in W2.
We tentatively conclude that the remaining members later than M8 do
not have excesses in [4.5] and W2.

To refine our identifications of excesses, we have checked
for sources that are redder than the boundaries for [4.5] or W2 in
Figure~\ref{fig:pan} but have data at longer wavelengths that are
consistent with a stellar photosphere. All stars of this kind are
only slightly above the excess thresholds in [4.5] and W2.
Therefore, we indicate that they lack [4.5] and W2 excesses in
Table~\ref{tab:all}. These stars consist of 
2MASS J16072682-1855239, 2MASS J16104202-2101319, 2MASS J16133688-2327298,
2MASS J16152024-2333588, 2MASS J16203456-2430205, 2MASS J16204144-2425491,
2MASS J16370753-2432395, Sco PMS 48, HIP~78977, Sco~PMS~42b, and HIP~81455.
The latter two stars may have small excesses in 
[24] (Section~\ref{sec:exc24}), but they lack excess emission in W3.
Several stars with excesses in [4.5] or W2 do not have photometry at
longer wavelengths, and thus cannot be examined in this way, because
they were below the detection limit (W3, W4, [24]) or were not observed
([5.8], [8.0], [24]).

We also have checked for members that have excesses in [4.5] but not W2,
or vice versa. Two objects of this kind
were discussed earlier in this section in the context of the
late-type members. Three additional stars show this discrepancy,
consisting of 2MASS J16090075-1908526, 2MASS J16095933-1800090,
and RX J1604.3-2130A.
As in the earlier cases, variability is the likely cause of the
differences between $K_s-[4.5]$ and $K_s-W2$ for these stars given
that other bands indicate the presence of disks.
For RX J1604.3-2130A, previous IR data have exhibited significant
variability as well \citep{dah09}.
However, the variability in this star is rather unusual.
Its $[4.5]-[8.0]$ color and its spectrum from \citet{dah09}
shortward 16~\micron\ are consistent with a stellar photosphere, but
the \wise bands over this same wavelength range (W1, W2, W3) show
excess emission relative to $K_s$ and each other, which may be caused by
changes in the structure of the inner disk \citep{esp11}.

\subsection{Excesses in [8.0]}

Photometry in [8.0] has been measured for 335 members of Upper Sco.
2MASS J16075850-2039485 is excluded from the diagram of $K_s-[8.0]$ in
Figure~\ref{fig:pan} because its 8~\micron\ flux is uncertain.
As done for $K_s-[4.5]$ and $K_s-W2$, we have defined a threshold
that varies with spectral type for identifying excesses in $K_s-[8.0]$.
It is defined by lines connecting (B0, 0.2), (K0, 0.32), (M0, 0.45), and
(M9, 1.01). Excesses from disks are larger at longer wavelengths, resulting in
a greater separation between diskless and disk-bearing stars in $K_s-[8.0]$
compared to $K_s-[4.5]$ and $K_s-W2$.
In all three of these colors, some of the early-type stars are 0.1--0.2~mag
redder than the sequence of the bluest stars.  Most of these excesses can be
attributed to extinction based on their lack of excesses at longer wavelengths
and their red near-IR colors. However, a few of these moderately red stars
may have small amounts of emission from debris disks \citep{car06}, even
though they appear below our adopted boundaries for excesses in
Figure~\ref{fig:pan}. Finally, we have checked for discrepancies between
apparent excesses in $K_s-[8.0]$ and data in other bands.
2MASS J16210222-2358395 is slightly above our
excess threshold for $K_s-[8.0]$, but [8.0] does not show an excess
relative to IRAC and WISE data at shorter wavelengths. Therefore, the
excess in $K_s-[8.0]$ may be due to stellar variability between the
observations at $K_s$ and the mid-IR bands.

\subsection{Excesses in W3}

In W3, W4, and [24], some members of Upper Sco are not detected or
have large uncertainties in their photometry.
For identifying excesses in these bands, we consider only measurements with
uncertainties less than 0.25~mag. We also exclude stars that are flagged
as extended. These criteria are satisfied in W3 by 658 members.
Our adopted threshold for excesses in $K_s-W3$ is shown in Figure~\ref{fig:pan},
which is defined by lines connecting (B0, 0.18), (G8, 0.33), and (M9, 1.52).

For stars that are slightly above the boundary for excesses, we have checked
whether excess emission is present in W4 or [24]. 
HIP~78977 lacks an excess in either band and 2MASS J16233234-2523485
should have little if any excess in W4 based on its non-detection.
Therefore, we conclude that these stars probably do not have excesses in W3.
Given the limits available at [24] and W4, 2MASS J16181618-2619080,
2MASS J16253672-2224285, 2MASS J16053077-2246200, and 2MASS J16044303-2318258
have colors of $\lesssim3$ in $K_s-W4$ and $K_s-[24]$, indicating that
they could have excess emission at W4 and [24]. Since we cannot rule out
the presence of excesses at longer wavelengths, we tentatively classify these
stars as having excesses in W3.

\citet{riz12} found that HIP~79878 has excess emission in W3, but it is
slightly below our threshold for $K_s-W3$. 
\citet{ria12} reported an excess in W3 for 2MASS J16185037-2424319,
but we rejected this measurement as a false detection.

\subsection{Excesses in [24] and W4}
\label{sec:exc24}

We discuss the bands [24] and W4 together since they have similar effective
wavelengths (23.7 and 22~\micron).
As with W3, we consider only data with uncertainties less than 0.25~mag
and that are not contaminated by extended emission.
The resulting sample contains 348 and 260 stars with photometry in [24] and W4,
respectively. Data are available in at least one of the two bands for 428
members.

Among the {\it Spitzer} and \wise bands that we are considering,
[24] and W4 are the only ones in which debris disks are likely to produce
noticeable excess emission (Section~\ref{sec:class}). Since these excesses
can be arbitrarily small, we have defined thresholds
in $K_s-[24]$ and $K_s-W4$ that identify the smallest
excesses that appear to be significant.
The threshold adopted for $K_s-W4$ is higher than that for $K_s-[24]$ since
the photometric uncertainties are larger in W4 than in [24], as illustrated
by the larger width in the sequence of photospheric colors for
$K_s-W4$ in Figure~\ref{fig:pan}. As a result, some stars that show
small excesses in $K_s-[24]$ are below the threshold for $K_s-W4$.
The thresholds are defined by (B0, 0.11), (K6, 0.57), and
(M9, 1.4) for $K_s-[24]$ and (B0, 0.34) and (M8, 1.15) for $K_s-W4$.

Most of our identifications of excesses in [24] and W4 agree with those
from previous studies \citep{car06,chen06,chen11,cie07,ria09,bow11,riz12,rom12},
with the following exceptions.
Our [24] photometry for HIP~81455 is 0.4~mag brighter than that
measured by \citet{chen11} with the same data. As a result, we find an
an excess in this band while \citet{chen11} did not.
This star is partially resolved from another fainter star at a separation
of $8\arcsec$, which was removed through PSF subtraction prior to our aperture
photometry.  We speculate that the PSF photometry of \citet{chen11} may
have been affected by this second star.
\citet{chen11} found that HIP~78977 has an excess in $K_s-[24]$,
but it is slightly below our thresholds for both $K_s-[24]$ and $K_s-W4$.
Neither [24] nor W4 show an excess relative to W1, W2, and W3
while excesses do appear in $K_s-W2$ and $K_s-W3$. These discrepant colors
are likely caused by variability of the star between the near- and mid-IR
observations, and we conclude that it does not have excesses in [24] and W4.
The following stars are above our [24] thresholds but slightly below
those adopted by \citet{car09}: HIP~78702, HIP~78099, HIP~76633, HIP~79250,
and RX~J1602.8-2401B.
\citet{riz12} used W4 to assess the presence of excess emission for
HIP~78265, HIP~80019, HIP~80311, and HIP~82319, but we have rejected
these data because they are false detections or too uncertain.

We have identified 22 new candidates in Upper Sco for weak excess
emission in [24] or W4, which may indicate the presence of debris disks
or evolved transitional disks (see Section~\ref{sec:class}).
Six of these stars were found to lack excesses in previous studies,
as described above. The remaining new candidates for weak excesses consist
of HD~142506, HIP~80196, 2MASS J16020287-2236139, 2MASS J16050231-1941554,
2MASS J16052459-1954419, 2MASS J16061330-2212537, 2MASS J16080555-2218070,
2MASS J16094098-2217594, 2MASS J16103956-1916524, 2MASS J16104202-2101319,
2MASS J16114612-1907429, 2MASS J16124893-1800525, 2MASS J16230783-2300596,
RXJ1614.6-1857, Sco~PMS~8b, and Sco~PMS~42b.
The detection of a W4 excess for HD~142506 is tentative since the photometry
may be affected by blending with another star.
Similarly, 2MASS J16235695-3312033 is above the threshold for an excess in
$K_s-W4$, but it is blended with another star. Because it lacks an excess in
the higher-resolution images at [24], we classify this star as lacking an
excess in W4.

Unlike the other {\it Spitzer} bands that we consider, the [24] images
did not detect all of the members of Upper Sco that they encompass.
Among the members observed in [24], 94 stars were not detected or have
errors large enough to be excluded from our analysis.
Three of these stars have early spectral types and lack data because
they are projected against bright nebulosity or are dominated by
extended emission. The remaining 91 objects have M types, most of which are
later than M3. A few of these stars have poor constraints on [24] because
of bright nebulosity or the PSF of another star, but most
have limits of $[24]\sim10$, indicating that these stars
have colors of $K_s-24\lesssim3.5$.
The detection limit from W4 is bright enough ($W4\sim8$) that the resulting
limits on $K_s-W4$ do not place useful constraints on the presence of disks
for most members of Upper Sco that are not detected in this band.

\section{Classification of Disks}
\label{sec:class}

Multiple stages of evolution are usually present among the circumstellar
disks in young stellar populations ($\tau\lesssim10$~Myr).
The evolution of a disk is largely characterized by a change in its
structure, which is reflected in its IR spectral energy distribution (SED).
As a result, one can use multi-band IR data like those we have
compiled from {\it Spitzer} and \wise for Upper Sco to estimate the
evolutionary stages of disks. In this section, we describe our adopted
definitions for these stages and apply them to Upper Sco.

\subsection{Terminology}

A variety of names and definitions have
been devised for the evolutionary stages of disks.
We use the ones adopted by \citet{esp12}, which are as follows:
A {\it full disk} is optically thick at IR wavelengths and
has not experienced significant clearing as indicated by its SED.
A {\it pre-transitional disk} and a {\it transitional disk} have
gaps and holes in their dust distributions, respectively, that are large
enough to noticeably affect their SEDs \citep[$\Delta R_{gap}>5$~AU,
$R_{hole}>1$~AU,][]{str89,cal05,esp07,esp10}.
An {\it evolved disk} is becoming optically thin and does not have a large gap
or hole \citep{her07}. Evolved disks also have been named anemic, thin, weak,
and homologously depleted \citep{lada06,bar07,dah07,cur09}.
Since we are examining a stellar population that includes relatively
advanced stages of disk evolution, we also use the terms
{\it evolved transitional disk}, which has optically thin primordial dust
and a large inner hole \citep{luh10tau}, and {\it debris disk},
which consists of second-generation dust that is generated by collisions
among planetesimals \citep{ken05,rie05}. All of these stages
except debris disks are considered {\it primordial disks}.
This definition of a primordial disk differs from the one that we adopted in
\citet{luh10tau}, where was equivalent to a full disk as defined here.

\subsection{Observational Criteria}
\label{sec:criteria}

To define our observational criteria for the different evolutionary stages
of disks, we begin by describing the qualitative features of their SEDs.
A full disk produces moderate-to-strong emission throughout the IR regime.
Because it has a large inner hole, a transitional disk is faint at
shorter IR wavelengths and is much brighter at longer wavelengths.
The emission from an evolved disk does not change abruptly with wavelength
since it lacks a large hole or gap, and it is very weak since the disk
is becoming optically thin.
An evolved transitional disk and a debris disk have similar SEDs,
showing only weak emission at long IR wavelengths.

To develop quantitative criteria for classifying disks, one approach
has been to identify distinct populations in the IR SEDs of young stars
and associate them with different evolutionary stages of disks.
Populations of this kind were first found in Taurus, where
most stars have mid-IR colors that are either consistent with stellar
photospheres, or are much redder. The few sources within the gap between
these two groups were described as ``in transition" \citep{str89,skr90}.
Similar gaps are present in the mid-IR colors of other star-forming regions
as well \citep[][references therein]{luh10tau}.
In some studies, the red edge of the gap in colors like $K_s-[8.0]$
has been adopted as the boundary between full disks and pre-transitional
disks on the redward side, and evolved and transitional disks
on the blueward side \citep{her07,luh10tau}.
This threshold seems reasonable given that it roughly coincides with the colors
produced by models of full disks with high degrees of dust settling
\citep{luh10tau,esp12}.
Distinctive populations of colors and supporting modeling also have been used
to guide the identification of debris disks and evolved transitional disks
\citep{car09}.

A different approach to classifying disks has been employed by
\citet{cur11}. They defined transitional disks to include disks with inner
holes or gaps and disks that are optically thin, which correspond to
transitional, pre-transitional, and evolved disks in our terminology.
To identify disks of this kind, \citet{cur11} compared observed SEDs
to the SEDs predicted by disk models. When millimeter measurements were
available, they also required that a disk have an estimated mass less than
0.001~$M_*$ in order for it to be considered optically thin.
After applying these procedures, \citet{cur11} concluded that some of the disks
previously classified as settled, optically thick disks (i.e., redward of the
gap in IR colors described above) are instead optically thin,
and hence qualify as transitional disks as defined in their study.
Because their definition of transitional disks was broader and their
classification criteria encompassed redder disks, \citet{cur11} found a higher
abundance of transitional disks in regions like Taurus than some studies
\citep[e.g.,][]{luh10tau}.
Of course, a comparison of the abundances of transitional disks,
and the resulting evolutionary timescales (Section~\ref{sec:histo}),
is not meaningful when the adopted definitions of transitional disks differ.

The classification scheme from \citet{cur11} has disadvantages.
Transitional disks are not defined in the traditional manner (disks with
large holes), which can lead to confusion when classifications from
different studies are compared. The observational criteria from
\citet{cur11} rely on theoretical SEDs, which are subject
to imperfections in the physical prescriptions of the models and
degeneracies among the disk parameters. The predicted SEDs also can
vary from one set of adopted models to another. For instance, as indicated
above, some objects can be reproduced with either an optically thin disk 
or a settled optically thick disk, depending on the choice of models
\citep{luh10tau,cur11,esp12}.
The use of disk mass as a criterion for identifying optically thin disks
is also problematic. The requisite millimeter data are 
unavailable for many members of young clusters and are
incomplete at low disk masses, even in Taurus \citep{and05}.
Estimates of disk masses and optical depths are sensitive to both the disk
model and the opacities that are adopted \citep{esp12}.
In addition, \citet{cur11} selected a threshold of 0.001~$M_*$ as a lower
mass limit for optically thick disks because nearly all optically thick disks
in Taurus with millimeter data have masses above this value.
However, 0.001~$M_*$ is only slightly above the detection limits for the
millimeter observations in Taurus \citep[$\gtrsim0.0005$~$M_\odot$,][]{and05},
and hence does not represent a measurement of the minimum mass of optically
thick disks.
Deeper millimeter data are needed to reliably determine the distribution
of disks masses down to low optical depths.

In a critique of the disk classifications based on colors from
\citet{luh10tau}, \citet{cur11} noted that UX~Tau~A has a disk with a large
gap \citep{esp07} but did not satisfy the color criteria for a transitional 
disk from \citet{luh10tau} and instead was classified as a primordial disk
(or a full disk with our current terminology).
However, this classification was consistent with the definitions adopted by
\citet{luh10tau} in which a disk with a gap was considered pre-transitional
and a subset of primordial disks, whereas only disks with large holes
were named transitional.
Any set of classification criteria will exclude disks with holes or gaps
below a certain size, so the absence of disks with gaps like UX~Tau~A from
our definition of transitional disk is not particularly worrisome.
We simply use the term ``transitional disk" to refer to a different range of
structural evolution than \citet{cur11}.

In summary, we have selected a classification scheme for disks in which the
evolutionary stages are defined by distinct structural characteristics
\citep{esp12}. The criteria for identifying members of these classes are based
on patterns in the data for disks (e.g., gap in colors) and can be applied to
data that are available for large numbers of young stars and brown dwarfs,
namely multi-band mid-IR photometry from {\it Spitzer} and {\it WISE}.

\subsection{Disk Classes in Upper Sco}

\begin{figure}
\epsscale{1}
\plotone{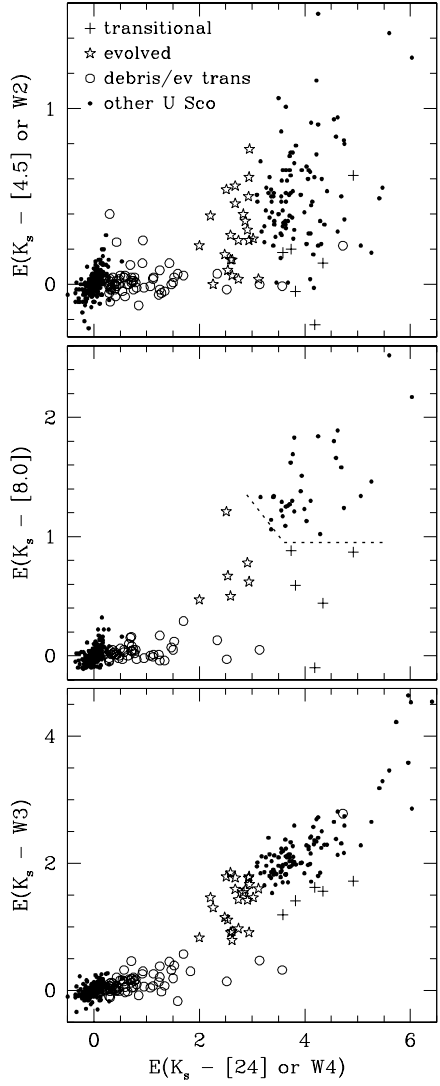}
\caption{
Infrared color excesses for members of Upper Sco. Data at 4.5 and
24~\micron\ from {\it Spitzer} are shown when available. Otherwise, 
measurements at similar wavelengths from \wise are used (W2 and W4).
We have indicated candidates for transitional disks (crosses), evolved disks
(stars), and debris or evolved transitional disks (circles).
In the middle diagram, we have marked the lower boundary that we have adopted
for full disks (dotted line).
}
\label{fig:ex}
\end{figure}

In \citet{luh10tau}, we used extinction-corrected mid-IR colors to estimate
the evolutionary stages of disks in Taurus and other regions.
For Upper Sco, we modify this approach by considering instead the color
excesses relative to photospheric colors (e.g., $E(K_s-[24]$). 
We have computed the excess in a given color for each member of Upper Sco
by subtracting the color of a stellar photosphere in Upper Sco for the
spectral type in question, as determined by a fit to the sequence of diskless
stars in Figure~\ref{fig:pan}. Through this process, the colors for
all members are corrected for roughly the same level of
extinction, which is the average extinction of the diskless members.
The resulting excesses for the colors from Figure~\ref{fig:pan}
are presented in Figure~\ref{fig:ex}. The Be stars are omitted
since they do not require classification.
We plot the excesses based on [4.5] and W2 together
because of the similarity of these bands. We do the same for [24] and W4.
For each object, we show the data in [4.5] and [24] when available, and
otherwise use W2 and W4.

\citet{luh10tau} used $K_s-[5.8]$, $K_s-[8.0]$, and $K_s-[24]$ for
their disk classifications. Since relatively few members of Upper Sco
have been observed in [5.8], we begin our classifications with
the other two colors (and $K_s-W4$). In Figure~\ref{fig:ex}, we show
our adopted thresholds in $E(K_s-[8.0])$ and $E(K_s-[24]/W4)$
for distinguishing between full disks and later evolutionary stages.
These thresholds approximate those applied to $K_s-[8.0]$ and $K_s-[24]$
by \citet{luh10tau}.
We have used the objects classified in this way to reveal the boundary
between full disks and later stages in $E(K_s-W3)$ vs.\ $E(K_s-[24]/W4)$, and
we have used this boundary to classify stars that lack data in [8.0].
To classify the disks bluer than full disks, we apply the following criteria,
which are similar to those from \citet{car09} and \citet{luh10tau}:
transitional disks have $E(K_s-[24]/W4)>3.2$;
evolved disks have $E(K_s-[24]/W4)<3.2$, $E(K_s-[8.0])>0.3$,
and $E(K_s-W3)>0.5$;
debris disks and evolved transitional disks have $E(K_s-[24]/W4)<3.2$,
$E(K_s-[8.0])<0.3$, and $E(K_s-W3)<0.5$.
Measurements of gas content are necessary for distinguishing between the latter
two classes of disks since they exhibit similar mid-IR SEDs \citep{car09},
but such data are unavailable for most disks in Upper Sco.
For the early-type stars, the high abundance of these debris/evolved
transitional candidates
relative to full disks indicates that most are probably debris disks
since evolved transitional disks are expected to be short-lived
(Section~\ref{sec:histo}).
Two exceptions to the above criteria are HIP~79288 and HIP~79977.
Both stars are redder than our thresholds for debris/evolved transitional
disks, but we classify them as such based on previous studies 
\citep{syl00,hon04,smi08,chen06,chen11}.

We have attempted to classify the disks that lack photometry in [24]
or W4, and hence do not appear in Figure~\ref{fig:ex}.
For the disks of this kind that are detected in W3, the excesses in this
band are small enough to indicate that they are evolved or transitional.
Some of those disks are probably evolved rather than transitional based
on the available constraints on $E(K_s-[24]/W4)$.
Two of the disks without W3, W4, or [24] have data in [8.0], both of which
are full disks according to their values of $E(K_s-[8.0])$. 
The remaining disks that lack data in [24] or W4 are detected
only at $\lambda<5$~\micron, most of which are later than M6.
We classify them as full disks for the purposes of this work, although some
could be evolved disks.

In Table~\ref{tab:all}, we present our disk classifications for all
members of Upper Sco that show mid-IR excess emission.
Through our classifications, we arrive at 135 full disks, 28 evolved disks,
6 transitional disks, three disks that are evolved or transitional,
and 63 debris/evolved transitional disks.
Previous mid-IR studies have already identified many of these 
full disks \citep{car06,sch07,ria12}, evolved and transitional
disks \citep{ria09,dah09,dah10,luh10tau}, and debris/evolved
transitional disks \citep{cie07,car09,chen06,chen11,rom12}.
Our new candidates for transitional disks consist of 2MASS J16082733-2217292,
2MASS J16101888-2502325, and 2MASS J16151239-2420091.
It is likely that a number of full disks have inner holes
that are too small to be classified as transitional using
our criteria, such as 2MASS J16081566-2222199, which has strong excesses
in W3 and W4 but no excess in W2 (see upper panel of Figure~\ref{fig:ex}).
Most of the candidates for evolved disks are newly identified as such in
this work.
In Section~\ref{sec:exc24}, we listed our new candidates for debris/evolved
transitional disks and a few stars in this category that we classify
differently than other studies.
We note that HIP~81474 has been frequently treated as a Herbig Ae/Be star,
but it lacks a clear detection of line emission \citep{her05}. Therefore, it
is classified as debris/evolved transitional in this work.

\section{Statistical Properties of Excesses}
\label{sec:global}

\subsection{Excess Fractions}
\label{sec:fraction}

\begin{figure}
\epsscale{1}
\plotone{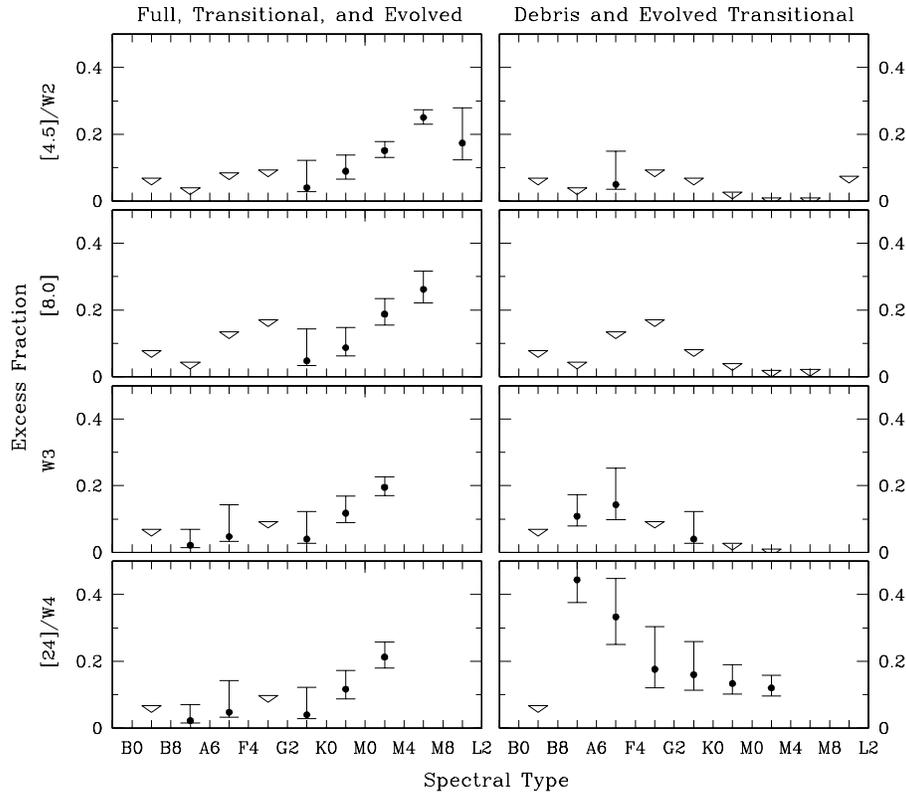}
\caption{
Excess fractions versus spectral type in Upper Sco
for full, transitional, and evolved disks (left) and debris and evolved
transitional disks (right, Table~\ref{tab:fex}).  For each band, data are shown
only down to the coolest spectral type at which most of the known members
are detected. The triangles represent 1~$\sigma$ upper limits.
}
\label{fig:fex}
\end{figure}

The distribution of disk excesses as a function of parameters like
wavelength, spectral type, cluster age, excess size, and disk class can
provide insight into the evolution of disks.
To characterize the distribution of excesses in Upper Sco, we begin by
considering the fraction of members that exhibit excesses.
In Table~\ref{tab:fex} and Figure~\ref{fig:fex}, we present
the excess fraction as a function of spectral
type for [4.5]/W2, [8.0], W3, and [24]/W4. To examine the evolution of
primordial and debris disks separately, one excess fraction is computed for
full, transitional, and evolved disks, and another is computed for debris
and evolved transitional disks. The latter disks do contain primordial dust,
but they cannot be distinguished from debris disks with available data, as
discussed in the previous section.

The images in W3, W4, and [24] were not sufficiently sensitive to detect
all known members of Upper Sco. The members that were too faint to be
detected tend to have later spectral types or smaller excesses.
As a result, we cannot compute meaningful excess fractions in these bands
at the latest types. Therefore, the excess fractions for W3 and [24]/W4
in Figure~\ref{fig:fex} are shown only for spectral types in which nearly
all of the known members are detected. 
For the joint [24]/W4 excess fraction, we adopt the [24] data when available,
which have a high level of completeness down to a spectral type of M4.
Among [24] non-detections earlier than M4, six stars were not subject
to extended emission or blending with other stars, and hence have
useful limits on their photometry. The resulting constraints on $K_s-[24]$
indicate that these six stars have little, if any, excess emission, so
we include them as non-excess stars in the [24]/W4 fraction.
Among stars that were not observed at [24], the W4 images detected all
but three stars earlier than M0. Therefore, we use the W4 data of members
not observed at [24] and earlier than M0
in the [24]/W4 fraction. The three stars earlier than M0 with non-detections
in W4 have sufficient constraints on this band to indicate that excesses
are probably not present, so we count them as non-excess stars.
It is unnecessary to impose such restrictions on the spectral types included
in the excess fractions at [4.5], W2, and [8.0] since all known 
members were detected in these bands as long as they were observed
(by IRAC) and not blended with another star.
The Be stars are excluded from these excess fractions.

We first discuss the excess fractions for the full, transitional, and evolved
disks (i.e., all primordial disks except for evolved transitional disks).
For these disks, the excess fractions are very low at a
roughly uniform level as a function of spectral type for B through G types,
and then increase with later types from K through M. This trend was discovered
by \citet{car06}, supported by additional data from \citet{sch07} for late-M
members, and now is reinforced further with better number statistics.
All of the bands in Figure~\ref{fig:fex} exhibit a similar dependence on
spectral type, except that the excess fractions are slightly higher at longer
wavelengths due to the presence of a small number of transitional disks.
As discussed by \citet{car06}, these data indicate that disk lifetimes
are longer for stars at lower masses, providing more time for planet formation.
We note that the excess fraction in [4.5]/W2 for the last bin in
Figure~\ref{fig:fex} at M8--L2 ($\sim0.01$~$M_\odot$) is uncertain because of
the small number of objects and the rapidly changing photospheric colors, 
which makes it difficult to reliably identify excesses.
In addition, since excesses in [4.5] are smaller at later types
\citep{luh10tau}, the excess fraction in this band may significantly
underestimate the disk fraction. Longer wavelengths are needed for more
robust detections of disk excesses for brown dwarfs, but the members
at M8--L2 were too faint for W3 and W4 photometry and were not observed at
[5.8] and [8.0].
The disk fractions based on the data in Figure~\ref{fig:fex} are
compared to similar measurements in younger clusters by \citet{luh12}.

We now consider the excess fractions for the debris and evolved transitional
disks. As mentioned earlier, we cannot distinguish between these two
types of disks with the mid-IR data, but most of the candidates at earlier
spectral types are probably debris disks. It is unclear which class of disk
dominates at later types. In Figure~\ref{fig:fex}, the excess fraction for
this sample increases rapidly with longer wavelengths. 
Excesses in [4.5]/W2 are present for only a few stars, which may be
some of the youngest debris disks in the association
\citep{syl00,hon04,smi08,chen06,chen11}.
The [24]/W4 fraction is highest for A stars ($\sim0.4$), drops sharply
for B stars ($<0.07$), and decreases more gradually with later types,
reaching 0.1--0.2 for G, K, and M stars.
These trends are consistent with those previously found in smaller
samples of [24] data in Upper Sco by \citet{car09} and \citet{chen11}.
Those studies investigated various aspects of their [24] data,
and their results apply to our new measurements as well.
For instance, \citet{car09} found that the rapid decrease in [24] fraction
from A to B stars is expected because of increased ejection of dust grains by
radiation pressure \citep[see also][]{ken08}.
The [24]/W4 excess fractions at later spectral types are more difficult to
interpret since they contain an unknown contribution from evolved transitional
disks. In addition, many debris disks around late-type stars may be too
cool to produce noticeable excesses in these bands.
\citet{car09} and \citet{chen11} also examined the evolution of debris disks
by comparing the [24] excess fractions among Upper Sco and other
associations and clusters across a range of ages
\citep[see also][]{sie07,cur08,gas09}. Our data have improved the statistical
constraints on excess fractions and sizes in Upper Sco, particularly at M
types, but an analysis like that from \citet{car09} and \citet{chen11}
for M stars is not possible because sufficiently complete samples of
[24] data are available for only a few regions.

\subsection{Sizes of Excesses at [24] and W4}
\label{sec:histo}

The mid-IR colors of the members of a given star-forming region
typically exhibit a gap between diskless stars and the bulk of the
disk-bearing stars (see Section~\ref{sec:criteria}).
This trough in the distribution of excess sizes has been previously interpreted
as evidence that once a disk has begun to clear to a certain degree,
the remainder of
the dust in the inner disk clears rapidly \citep{skr90}. In this section,
we examine the constraints on the clearing timescale that are provided
by the distribution of excesses in Upper Sco.

Because the contrast between a stellar photosphere and circumstellar disk 
increases at longer IR wavelengths, [24] and W4 offer the largest dynamic range
in the sizes of excesses among the photometric bands that we have compiled for
Upper Sco. Given that the [24] observations are more sensitive than those
at W4, we choose the former for constructing the distribution of excesses.
To characterize how this distribution changes with cluster age, we compare
the data for Upper Sco to measurements of [24] in Taurus \citep{luh10tau}.
Upper Sco and Taurus are the best available regions for such a comparison
since they roughly bracket the ages of primordial disks ($\sim1$ and 10~Myr),
contain two of the richest and nearest samples of young stars, and have been
thoroughly observed by {\it Spitzer}. In particular, few other
regions have [24] measurements that reach the stellar photospheres of low-mass
stars, which is essential for fully sampling excesses across all sizes. 
Since very few primordial disks are present among the early-type stars in
Upper Sco, we exclude stars earlier than K0 from the distributions of
$K_s-[24]$ excesses for Upper Sco and Taurus. We also omit stars later
than M5 since most of the diskless stars at these types
are below the detection limits of the [24] images. Roughly a dozen stars
in each region at K5--M5 are not detected in [24] and could have small
excesses based on the limits on $K_s-[24]$. As a result, the distributions
of excesses that we construct are biased slightly against small excesses.

\begin{figure}
\epsscale{1}
\plotone{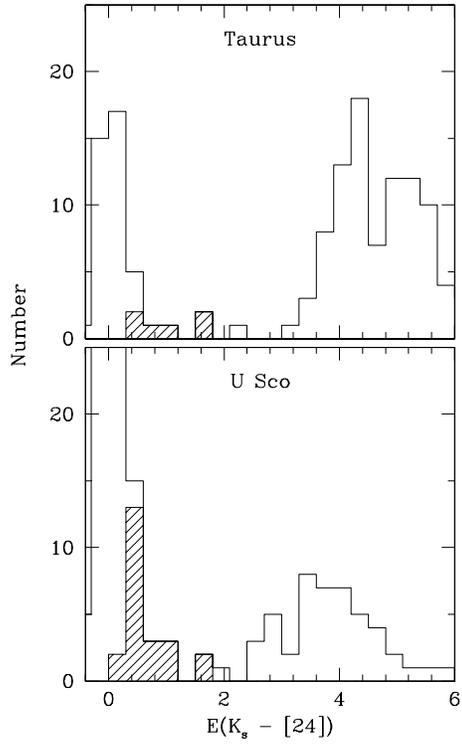}
\caption{
Distributions of E($K_s$-[24]) for K0--M5 members of Taurus and Upper Sco.
The shaded histograms represent stars that may have debris disks or
evolved transitional disks based on their small 24~\micron\ excesses.
Protostars have been omitted.
}
\label{fig:histo}
\end{figure}

The distributions of $E(K_s-[24])$ for K0--M5 members of Taurus and Upper Sco
are presented in Figure~\ref{fig:histo}, excluding protostellar sources.
The two distributions exhibit similar troughs at $E(K_s-[24])\sim0.5$--3.
Most observations of other young clusters at [24] are less complete
at low excesses or consider smaller numbers of stars, but they are generally
consistent with gaps of this kind \citep{luh10tau}.
Meanwhile, among stars redward of the trough ($E(K_s-[24])>3$), the average
color is bluer in Upper Sco than in Taurus. 
These characteristics can be explained in the following manner.
The youngest systems have highly flared disks, and hence the largest IR
excesses. As they evolve (e.g., growth of grains, accretion onto the star),
disks become more settled and produce smaller excesses. 
After disks have become highly settled and have reached a certain level of
initial clearing (corresponding to $E(K_s-[24])\sim3$), they quickly clear
their remaining
inner disks, resulting in a scarcity of disks with small excesses.
Because star formation is ongoing in Taurus but has ceased in Upper Sco,
the distribution of excesses in the former is more heavily weighted to younger
and redder disks.  In other words, the youngest disks in Upper Sco have already 
evolved substantially, explaining its paucity of very red disks relative to 
Taurus. Because the evolutionary stages of disks are not sampled in the same
way by the differing star formation histories of Taurus and Upper Sco, a
comparison of the N(evolved+transitional)/N(full) does not indicate the
relative timescales of clearing between the two populations \citep{luh10tau}.
Instead, one must compare the ratio of the number of evolved and transitional
disks to the number of sources slightly redward of the trough, i.e., the shape
of the distribution of excesses at $E(K_s-[24])\lesssim3.5$.
This property is not tightly constrained in either region because of the
small numbers of stars involved, but Taurus and Upper Sco do not show
any obvious differences. Thus, there is no evidence to indicate that
the clearing timescale changes from the age of Taurus to the age of Upper Sco
($\sim1$--10~Myr). The data in Figure~\ref{fig:histo} do reveal a
higher abundance of candidates for evolved transitional disks or debris disks
in Upper Sco. If the lifetime of the evolved transitional stage is similar
to that of the transitional and evolved stages, then most of these candidates
in Upper Sco are probably debris disks. It is expected that Upper Sco would
contain more debris disks than Taurus given their relative ages.

As discussed in Section~\ref{sec:criteria}, the definition of transitional
disks adopted by \citet{cur11} encompasses some of the objects redward
of the gap in the mid-IR colors in regions like Taurus and Upper Sco.
As a result, they have derived higher abundances of transitional disks
in young clusters, and hence a longer timescale for inner disk clearing, than
found here and in other studies that use a more restrictive definition
for transitional disks \citep{her07,luh10tau}. 
Of course, it is not meaningful to compare abundances of transitional disks
and the resulting timescales among studies that do not define this evolutionary
stage in the same manner.
We contend that the clearing timescales from \citet{cur11} are less useful
and practical since they are based on a definition of transitional disks
that is subject to uncertainties and degeneracies in model SEDs and that does
not rely on distinct observational signatures (i.e., gap in colors).

\citet{cur09} suggested that the gap in mid-IR colors in Taurus is not
evidence of a short lifetime for the evolutionary stage of disks within
this gap, contrary to the original interpretation \citep{skr90}.
Instead, they argued that the gap is present because few members of Taurus
are old enough to have reached that stage, and that the diskless stars
in Taurus have lost their disks due to binary companions rather than
advanced age. In \citet{luh10tau}, we noted that a rather large fraction
of the Taurus population is diskless (40\% among non-protostars) and that
these stars are older than disk-bearing members, on average, based on
their wider spatial distribution, demonstrating that a significant number of
stars in Taurus have aged sufficiently to fully clear their disks,
and hence pass through the evolved and transitional stages.
\citet{cur11} reiterated the same points from their previous study, but
also conceded that many of the diskless stars in Taurus may indeed be older
than their counterparts with disks.
Here, we once again contend that if a significant number of stars
have aged sufficiently to have fully cleared their disks, then the
evolved and transitional stages should be fully sampled as well,
and therefore their scarcity demonstrates that they are short-lived
relative to full disks.
This conclusion is supported by the fact that older regions like Upper
Sco exhibit the same gap in their colors
\citep[Figure~\ref{fig:histo};][references therein]{luh10tau}.

\section{Conclusions}

We have performed a census of the circumstellar disk population of the Upper
Sco association ($\tau\sim11$~Myr) using mid-IR images obtained with
the {\it Spitzer Space Telescope} and {\it WISE}.
The results of this study are summarized as follows:

\begin{enumerate}

\item
We have analyzed all images of known members of Upper Sco
at 3.6, 4.5, 5.8, 8.0, and 24~\micron\ that have been collected
by {\it Spitzer} to date. Among the 863 known members, 484 have been
observed in at least one of these bands. We have presented a catalog of
photometry and non-detections for these objects.

\item
We have compiled all photometry at 3.4, 4.6, 12, and 22~\micron\ (W1--W4) for
members of Upper Sco from the \wise mission, thoroughly vetting these data
for false detections, blends with neighboring stars, and extended emission. 
All resolved, unblended members were detected by \wise in at least 3.4 and
4.6~\micron.

\item
We have used colors computed between $K_s$ and six of the {\it Spitzer} and
\wise bands to identify the members that exhibit excess emission from
circumstellar disks. 
We have estimated the evolutionary stages of the detected disks using the
sizes of the excesses among these bands.
These stages are defined based on distinct structural characteristics
that are apparent in the SEDs \citep[][references therein]{esp12}, and they
consist of the following: 
optically thick disks without large holes or gaps ({\it full}),
optically thick disks with large holes ({\it transitional}),
optically thin disks without large holes or gaps ({\it evolved}),
optically thin disks with large holes ({\it evolved transitional}),
and disks of second-generation dust ({\it debris}).
Through these classifications, we have found $\sim50$ new candidates for
transitional, evolved, and debris disks.

\item
Based on our census of disks,
the fraction of Upper Sco members with inner primordial disks is
$\lesssim10$\% for B--G stars ($M>1.2$~$M_\odot$) and increases with later
types, reaching $\sim25$\% at M5--L0 ($M\sim0.01$--0.2~$M_\odot$).
This trend was first observed by \citet{car06}, and is now reinforced
with a sample that is larger and reaches lower stellar masses.
The combination of these data and the latest age estimate for Upper Sco
\citep[$\tau\sim11$~Myr,][]{pec12} indicate that a significant fraction
of primordial disks around low-mass stars survive for at least $\sim10$~Myr.

\item
K0--M5 members of Upper Sco show a dearth of disks with small excesses
(transitional and evolved disks).
When the differences in star formation histories of Upper Sco and Taurus
($\tau\sim1$~Myr) 
are taken into account, the distributions of excesses for both regions
indicate a timescale for inner disk clearing that is 
much shorter than the typical lifetime of primordial disks.

\end{enumerate}

\acknowledgements

We thank Catherine Espaillat, Nuria Calvet, and Eric Feigelson
for comments on the manuscript.
K. L. was supported by grants AST-0544588 and NNX12AI58G from the National
Science Foundation and the NASA Astrophysics Data Analysis Program,
respectively, and E. M. was supported by grant AST-1008908 from the National
Science Foundation.
This work makes use of data from the {\it Spitzer Space Telescope}, 2MASS,
{\it WISE}, and the Infrared Processing and Analysis Center (IPAC)
Infrared Science Archive (IRSA).
{\it Spitzer} and IRSA are operated by the Jet Propulsion Laboratory (JPL) and 
the California Institute of Technology (Caltech) under contract with NASA. 
\wise is a joint project of the University of California, Los Angeles,
and JPL/Caltech, funded by NASA. 2MASS is a joint project of the University of 
Massachusetts and the IPAC/Caltech, funded by NASA and the NSF.
The Center for Exoplanets and Habitable Worlds is supported by the
Pennsylvania State University, the Eberly College of Science, and the
Pennsylvania Space Grant Consortium.

\clearpage

\LongTables

\begin{deluxetable}{ll}
\tablewidth{5in}
\tablecaption{Data for Members of Upper Sco\label{tab:all}}
\tablehead{
\colhead{Column Label} &
\colhead{Description}}
\startdata
Name & Source name\tablenotemark{a} \\
OtherNames & Other source names\tablenotemark{b} \\
SpType & Spectral type \\
r\_SpType & Spectral type reference\tablenotemark{c} \\
Adopt & Adopted spectral type \\
3.6mag & {\it Spitzer} [3.6] band magnitude \\
e\_3.6mag & Error in [3.6] band magnitude \\
f\_3.6mag & Flag on [3.6] band magnitude\tablenotemark{d} \\
4.5mag & {\it Spitzer} [4.5] band magnitude \\
e\_4.5mag & Error in [4.5] band magnitude \\
f\_4.5mag & Flag on [4.5] band magnitude\tablenotemark{d} \\
5.8mag & {\it Spitzer} [5.8] band magnitude \\
e\_5.8mag & Error in [5.8] band magnitude \\
f\_5.8mag & Flag on [5.8] band magnitude\tablenotemark{d} \\
8.0mag & {\it Spitzer} [8.0] band magnitude \\
e\_8.0mag & Error in [8.0] band magnitude \\
f\_8.0mag & Flag on [8.0] band magnitude\tablenotemark{d} \\
24mag & {\it Spitzer} [24] band magnitude \\
e\_24mag & Error in [24] band magnitude \\
f\_24mag & Flag gon [24] band magnitude\tablenotemark{d} \\
W1mag & {\it WISE} W1 band magnitude\tablenotemark{e} \\
e\_W1mag & Error in W1 band magnitude \\
f\_W1mag & Flag on W1 band magnitude\tablenotemark{d} \\
W2mag & {\it WISE} W2 band magnitude\tablenotemark{e} \\
e\_W2mag & Error in W2 band magnitude \\
f\_W2mag & Flag on W2 band magnitude\tablenotemark{d} \\
W3mag & {\it WISE} W3 band magnitude\tablenotemark{e} \\
e\_W3mag & Error in W3 band magnitude \\
f\_W3mag & Flag on W3 band magnitude\tablenotemark{d} \\
W4mag & {\it WISE} W4 band magnitude\tablenotemark{e} \\
e\_W4mag & Error in W4 band magnitude \\
f\_W4mag & Flag on W4 band magnitude\tablenotemark{d} \\
Exc4.5 & Excess present in [4.5]?  \\
Exc8.0 & Excess present in [8.0]? \\
Exc24 & Excess present in [24]? \\
ExcW2 & Excess present in W2? \\
ExcW3 & Excess present in W3? \\
ExcW4 & Excess present in W4? \\
DiskType & Disk type \\
\enddata
\tablenotetext{a}{Coordinate-based identifications from the 2MASS Point Source
Catalog when available.  Otherwise, identifications from \citet{all07}
and \citet{lod07} are listed.}
\tablenotetext{b}{Designations from the Hipparcos Catalog \citep{van07} and the
Henry Draper Catalog and other frequently used names.}
\tablenotetext{c}{
(1) \citet{hou88};
(2) K. Luhman, in preparation;
(3) \citet{kun99};
(4) \citet{pre98};
(5) \citet{tor06};
(6) \citet{lod06};
(7) \citet{lod08};
(8) \citet{hou82};
(9) \citet{pec12};
(10) \citet{cor84};
(11) \citet{shk09};
(12) \citet{wal94};
(13) \citet{mar10};
(14) \citet{pre02};
(15) \citet{kra09};
(16) \citet{mar04};
(17) \citet{sle08};
(18) \citet{ard00};
(19) \citet{sle06};
(20) \citet{ria06};
(21) \citet{her09};
(22) \citet{pre01};
(23) \citet{bej08};
(24) \citet{kra07};
(25) \citet{lod11};
(26) \citet{bil11};
(27) \citet{pra03};
(28) \citet{eis05};
(29) \citet{giz02};
(30) \citet{pra07};
(31) \citet{mar98a};
(32) \citet{mar98b};
(33) \citet{cow69};
(34) \citet{luh05usco};
(35) \citet{lut77};
(36) \citet{bow11};
(37) \citet{rom12};
(38) \citet{jay06};
(39) \citet{clo07};
(40) \citet{luh07oph};
(41) E. Mamajek, unpublished;
(42) \citet{gar67};
(43) \citet{her05}.}
\tablenotetext{d}{
nodet = non-detection;
sat = saturated;
out = outside of the camera's field of view;
bl = photometry may be affected by blending with a nearby star;
ext = photometry is known or suspected to be contaminated by extended emission
(no data given when extended emission dominates);
bin = includes an unresolved binary companion;
unres = too close to a brighter star to be detected;
false = detection from {\it WISE} catalog that appears false or unreliable
based on visual inspection;
err = W2 magnitudes brighter than $\sim$6~mag are erroneous.}
\tablenotetext{e}{
W1 and W2 are from the All-Sky Source Catalog and
W3 and W4 are from the Preliminary Release Source Catalog.}
\tablecomments{The table is available in a machine-readable format.}
\end{deluxetable}

\clearpage

\begin{deluxetable}{llllll}
\tablecolumns{6}
\tabletypesize{\scriptsize}
\tablewidth{0pt}
\tablecaption{Excess Fractions in Upper Sco\label{tab:fex}}
\tablehead{
\colhead{Spectral Type} &
\colhead{Mass\tablenotemark{a}} &
\colhead{[4.5]/W2} &
\colhead{[8.0]} &
\colhead{W3} &
\colhead{[24]/W4} \\
\colhead{} &
\colhead{($M_\odot$)} &
\colhead{} &
\colhead{} &
\colhead{} &
\colhead{} 
}
\startdata
\cutinhead{Full, Transitional, and Evolved Disks}
B0--B8 &  2.8--18 & 0/25 = $<0.07$ & 0/22 = $<0.08$ & 0/25 = $<0.07$ & 0/26 = $<0.07$  \\
B8--A6 & 1.8--2.8 & 0/45 = $<0.04$ & 0/40 = $<0.04$ & 1/46 = $0.02^{+0.05}_{-0.01}$ & 1/45 = $0.02^{+0.05}_{-0.01}$  \\
A6--F4 & 1.5--1.8 & 0/20 = $<0.09$ & 0/12 = $<0.13$ & 1/21 = $0.05^{+0.09}_{-0.02}$ & 1/21 = $0.05^{+0.09}_{-0.02}$  \\
F4--G2 & 1.4--1.5 & 0/18 = $<0.09$ & 0/9 = $<0.17$ & 0/18 = $<0.09$ & 0/17 = $<0.10$  \\
G2--K0 & 1.3--1.4 & 1/25 = $0.04^{+0.08}_{-0.01}$ & 1/21 = $0.05^{+0.09}_{-0.02}$ & 1/25 = $0.04^{+0.08}_{-0.01}$ & 1/25 = $0.04^{+0.08}_{-0.01}$  \\
K0--M0 & 0.7--1.3 & 6/67 = $0.09^{+0.05}_{-0.02}$ & 4/46 = $0.09^{+0.06}_{-0.03}$ & 8/68 = $0.12^{+0.05}_{-0.03}$ & 7/60 = $0.12^{+0.05}_{-0.03}$  \\
M0--M4 & 0.2--0.7 & 35/231 = $0.15^{+0.03}_{-0.02}$ & 18/96 = $0.19^{+0.04}_{-0.04}$ & 41/210 = $0.20^{+0.03}_{-0.03}$ & 23/108 = $0.21^{+0.05}_{-0.03}$  \\
M4--M8 & 0.035--0.2 & 97/387 = $0.25^{+0.02}_{-0.02}$ & 22/84 = $0.26^{+0.06}_{-0.04}$ &      \nodata &      \nodata  \\
M8--L2 & 0.01--0.035 & 4/23 = $0.17^{+0.11}_{-0.05}$ &      \nodata &      \nodata &      \nodata  \\
\cutinhead{Debris and Evolved Transitional Disks}
B0--B8 &  2.8--18 & 0/25 = $<0.07$ & 0/22 = $<0.08$ & 0/25 = $<0.07$ & 0/26 = $<0.07$  \\
B8--A6 & 1.8--2.8 & 0/45 = $<0.04$ & 0/40 = $<0.04$ & 5/46 = $0.11^{+0.06}_{-0.03}$ & 20/45 = $0.44^{+0.08}_{-0.06}$  \\
A6--F4 & 1.5--1.8 & 1/20 = $0.05^{+0.10}_{-0.01}$ & 0/12 = $<0.13$ & 3/21 = $0.14^{+0.11}_{-0.04}$ & 7/21 = $0.33^{+0.12}_{-0.08}$  \\
F4--G2 & 1.4--1.5 & 0/18 = $<0.09$ & 0/9 = $<0.17$ & 0/18 = $<0.09$ & 3/17 = $0.18^{+0.12}_{-0.06}$  \\
G2--K0 & 1.3--1.4 & 0/25 = $<0.07$ & 0/21 = $<0.08$ & 1/25 = $0.04^{+0.08}_{-0.01}$ & 4/25 = $0.16^{+0.10}_{-0.05}$  \\
K0--M0 & 0.7--1.3 & 0/67 = $<0.03$ & 0/46 = $<0.04$ & 0/68 = $<0.03$ & 8/60 = $0.13^{+0.06}_{-0.03}$  \\
M0--M4 & 0.2--0.7 & 0/231 = $<0.01$ & 0/96 = $<0.02$ & 0/210 = $<0.01$ & 13/108 = $0.12^{+0.04}_{-0.02}$  \\
M4--M8 & 0.035--0.2 & 0/387 = $<0.01$ & 0/84 = $<0.02$ &      \nodata &      \nodata  \\
M8--L2 & 0.01--0.035 & 0/23 = $<0.07$ &      \nodata &      \nodata &      \nodata  \\
\enddata
\tablenotetext{a}{Masses that correspond to the given range of spectral types
for an age of 11~Myr \citep{bar98,cha00,pal99,dot08,pec12}.}
\end{deluxetable}

\clearpage

\begin{thebibliography}{}

\bibitem[Allers et al.(2007)]{all07}
Allers, K. N., Jaffe, D. T., Luhman, K. L., et al. 2007, \apj, 657, 511

\bibitem[Andrews \& Williams(2005)]{and05}
Andrews, S. M., \& Williams, J. P. 2005, \apj, 631, 1134

\bibitem[Ardila et al.(2000)]{ard00}
Ardila, D., Mart{\'\i}n, E. L., \& Basri, G. 2000, \aj, 120, 479

\bibitem[Baraffe et al.(1998)]{bar98}
Baraffe, I., Chabrier, G., Allard, F., \& Hauschildt, P. H. 1998, \aap, 337, 403

\bibitem[Barrado y Navascu\'es et al.(2007)]{bar07}
Barrado y Navascu\'es, D., Stauffer, J. R., Morales-Calder\'{o}n, M., et al.
2007, \apj, 664, 481

\bibitem[Bejar et al.(2008)]{bej08}
B\'{e}jar, V. J. S., Zapatero Osorio, M. R., P\'{e}rez-Garrido, A., et al.
2008, \apj, 673, L185

\bibitem[Biller et al.(2011)]{bil11}
Biller, B., Allers, K., Liu, M., Close, L. M., \& Dupey, T. 2011, \aj, 730, 39

\bibitem[Bowler et.(2011)]{bow11}
Bowler, B. P., Liu, M. C., Kraus, A. L., Mann, A. W., \& Ireland, M. J. 2011,
\apj, 743, 148


\bibitem[Calvet et al.(2005)]{cal05}
Calvet, N., D'Alessio, P., Watson, D. M., et al. 2005, \apj, 630, L185

\bibitem[Carpenter et al.(2006)]{car06}
Carpenter, J. M., Mamajek, E. E., Hillenbrand, L. A., \& Meyer, M. R. 2006,
\apj, 651, L49

\bibitem[Carpenter et al.(2008)]{car08}
Carpenter, J. M., Bouwman, J., Silverstone, M. D., et al. 2008, \apjs, 179, 423

\bibitem[Carpenter et al.(2009)]{car09}
Carpenter, J. M., Mamajek, E. E., Hillenbrand, L. A., \& Meyer, M. R. 2009,
\apj, 705, 1646

\bibitem[Casali et al.(2007)]{cas07}
Casali, M., Adamson, A., Alves de Oliveira, C., et al. 2007, \aap, 467, 777

\bibitem[Chabrier et al.(2000)]{cha00}
Chabrier, G., Baraffe, I., Allard, F., \& Hauschildt, P. 2000, \apj, 542, 464

\bibitem[Chen et al.(2005)]{chen05}
Chen, C. H., Jura, M., Gordon, K. D., \& Blaylock, M. 2005, \apj, 623, 493

\bibitem[Chen et al.(2011)]{chen11}
Chen, C. H., Mamajek, E. E., Bitner, M. A., et al. 2011, \apj, 738, 122

\bibitem[Chen et al.(2006)]{chen06}
Chen, C. H., Sargent, B. A., Bohac, C., et al. 2006, \apjs, 166, 351

\bibitem[Cieza et al.(2007)]{cie07}
Cieza, L., Padgett, D. L., Stapelfeldt, K. R., et al. 2007, \apj, 667, 308

\bibitem[Cieza et al.(2010)]{cie10}
Cieza, L. A., Schreiber, M. R., Romero, G. A., et al. 2010, \apj, 712, 925

\bibitem[Close et al.(2007)]{clo07}
Close, L. M., Zuckerman, B., Song, I., et al. 2007, \apj, 660, 1492

\bibitem[Corbally(1984)]{cor84}
Corbally, C. J. 1984, \apjs, 55, 657

\bibitem[Cot\'e \& van Kerkwijk(1993)]{cot93}
Cot\'e, J., \& van Kerkwijk, M. H. 1993, \aap, 274, 870

\bibitem[Cowley et al.(1969)]{cow69}
Cowley, A., Cowley, C., Jaschek, M., \& Jaschek, C. 1969, \aj, 74, 375

\bibitem[Crampton(1968)]{cra68}
Crampton, D. 1968, \aj, 73, 338

\bibitem[Currie et al.(2008)]{cur08}
Currie, T., Plavchan, P., \& Kenyon, S. J. 2008, \apj, 688, 597

\bibitem[Currie et al.(2009)]{cur09}
Currie, T., Lada, C. J., Plavchan, P., Robitaille, T. P., Irwin, J., \&
Kenyon, S. J. 2009, \apj, 698, 1

\bibitem[Currie \& Sicilia-Aguilar(2011)]{cur11}
Currie, T., \& Sicilia-Aguilar, A. 2011, \apj, 732, 24


\bibitem[Cutri et al.(2012)]{cut12}
Cutri, R. M., Wright, E. L., Conrow, T., et al. 2012,
Explanatory Supplement to the WISE All-Sky Data Release Products

\bibitem[Dahm(2010)]{dah10}
Dahm, S. E. 2010, \aj, 140, 1444

\bibitem[Dahm \& Hillenbrand(2007)]{dah07}
Dahm, S. E., \& Hillenbrand, L. A. 2007, \aj, 133, 2072

\bibitem[Dahm \& Carpenter(2009)]{dah09}
Dahm, S. E., \& Carpenter, J. M. 2009, \aj, 137, 4024


\bibitem[Dotter et al.(2008)]{dot08}
Dotter, A., Chaboyer, B., Jevremovi\'{c}, D., et al. 2008, \apjs, 178, 89

\bibitem[de Geus et al.(1989)]{deg89}
de Geus, E. J., de Zeeuw, P. T., \& Lub, J. 1989, \aap, 216, 44

\bibitem[Eisner et al.(2005)]{eis05}
Eisner, J. A., Hillenbrand, L. A., White, R. J., Akeson, R. L., \& Sargent,
A. I. \apj, 623, 952

\bibitem[Espaillat et al.(2007)]{esp07}
Espaillat, C., Calvet, N., D'Alessio, P., et al. 2007, \apj, 670, L135

\bibitem[Espaillat et al.(2010)]{esp10}
Espaillat, C., D'Alessio, P., Hern\'{a}ndez, J., et al. 2010, \apj, 717, 441

\bibitem[Espaillat et al.(2011)]{esp11}
Espaillat, C., Furlan, E., D'Alessio, P., et al. 2011, \apj, 728, 49

\bibitem[Espaillat et al.(2012)]{esp12}
Espaillat, C., Ingleby, L., Hern\'{a}ndez, J., et al. 2012, \apj, 747, 103

\bibitem[Evans et al.(2009)]{eva09}
Evans, N. J., II, Dunham, M. M., J{\o}rgensen, J. K., et al. 2009, \apjs, 181, 321

\bibitem[Fazio et al.(2004)]{faz04}
Fazio, G. G., Hora, J. L., Allen, L. E., et al. 2004, \apjs, 154, 10

\bibitem[Garrison(1967)]{gar67}
Garrison, R. F. 1967, \apj, 147, 1003

\bibitem[Gaspar et al.(2009)]{gas09}
Gaspar, A., Rieke, G. H., Su, K. Y. L., et al. 2009, \apj, 697, 1578

\bibitem[Gizis(2002)]{giz02}
Gizis, J. E. 2002, \apj, 575, 484

\bibitem[Gutermuth et al.(2009)]{gut09}
Gutermuth, R. A., Megeath, S. T., Myers, P. C., et al. 2009, \apjs, 184, 18

\bibitem[Hambly et al.(2008)]{ham08}
Hambly, N. C., Collins, R. S., Cross, N. J. G., et al. 2008, \mnras, 384, 637

\bibitem[Herczeg et al.(2009)]{her09}
Herczeg, G. J., Cruz, K. L., \& Hillenbrand, L. A. 2009, \apj, 696, 1589

\bibitem[Hern\'{a}ndez et al.(2005)]{her05}
Hern\'{a}ndez, J., Calvet, N., Hartmann, L., et al. 2005, \aj, 129, 856

\bibitem[Hern\'andez et al.(2007)]{her07}
Hern\'andez, J., Hartmann, L., Megeath, T., et al. 2007, \apj, 662, 1067

\bibitem[Hewett et al.(2006)]{hew06}
Hewett, P. C., Warren S. J., Leggett S. K., \& Hodgkin S. L., 2006, \mnras,
367, 545

\bibitem[Hodgkin et al.(2009)]{hod09}
Hodgkin, S. T., Irwin, M. J., Hewett, P. C., \& Warren, S. J. 2009, \mnras,
394, 675

\bibitem[Honda et al.(2004)]{hon04}
Honda, M., Kataza, H., Okamoto, Y. K., et al. 2004, \apj, 610, L49

\bibitem[Houk(1982)]{hou82}
Houk, N. 1982, Catalogue of Two-dimensional Spectral Types for the HD Stars.
Vol. 3, (Ann Arbor: Univ. Mich.)

\bibitem[Houk \& Smith-Moore(1988)]{hou88}
Houk, N., \& Smith-Moore, M. 1988, Michigan Catalogue of Two-dimensional
Spectral Types for the HD Stars. Vol. 4, (Ann Arbor: Univ. Mich.)

\bibitem[Ireland et al.(2011)]{ire11}
Ireland, M. J., Kraus, A., Martinache, F., Law, N., \& Hillenbrand, L. A. 2011,
\apj, 726, 113

\bibitem[Jaschek et al.(1964)]{jas64}
Jaschek, C., Jaschek, M., \& Kucewicz, B. 1964, Z. Astrophys., 59, 108

\bibitem[Jayawardhana \& Ivanov(2006)]{jay06}
Jayawardhana, R., \& Ivanov, V. D. 2006, Science, 313, 1279

\bibitem[Kenyon \& Bromley(2005)]{ken05}
Kenyon, S. J., \& Bromley, B. C. 2005, \apj, 130, 269

\bibitem[Kenyon \& Bromley(2008)]{ken08}
Kenyon, S. J., \& Bromley, B. C. 2008, \apjs, 179, 451

\bibitem[Kimeswenger et al.(2004)]{kim04}
Kimeswenger, S., Lederle, C., Richichi, A., et al. 2004, \aap, 413, 1037

\bibitem[Kraus \& Hillenbrand(2007)]{kra07}
Kraus, A. L., \& Hillenbrand, L. A. 2007, \apj, 664, 1167

\bibitem[Kraus \& Hillenbrand(2009)]{kra09}
Kraus, A. L., \& Hillenbrand, L. A. 2009, \apj, 703, 1511

\bibitem[Kraus et al.(2005)]{kra05}
Kraus, A. L., White, R. J., \& Hillenbrand, L. A. 2005, \apj, 633, 452

\bibitem[Kunkel(1999)]{kun99}
Kunkel, M. 1999, Ph.D. thesis, Julius-Maximilians-Univ., Wu\"{u}rzburg

\bibitem[Lada et al.(2006)]{lada06}
Lada, C. J., Muench, A. A., Luhman, K. L., et al. 2006, \aj, 131, 1574

\bibitem[Lafreni{\`e}re et al.(2011)]{laf11}
Lafreni{\`e}re, D., Jayawardhana, R., Janson, M., et al. 2011, \apj, 730, 42

\bibitem[Lafreni{\`e}re et al.(2008)]{laf08}
Lafreni{\`e}re, D., Jayawardhana, R., \& van Kerkwijk, M. H. 2008, \apj, 689,
L153

\bibitem[Lawrence et al.(2007)]{law07}
Lawrence, A., Warren, S. J., Almaini, O., et al. 2007, \mnras, 379, 1599

\bibitem[Lodieu et al.(2011)]{lod11}
Lodieu, N., Dobbie, P. D., \& Hambly, N. C. 2011, \aap, 527, A24

\bibitem[Lodieu et al.(2006)]{lod06}
Lodieu, N., Hambly, N. C., \& Jameson, R. F. 2006, \mnras, 373, 95

\bibitem[Lodieu et al.(2008)]{lod08}
Lodieu, N., Hambly, N. C., Jameson, R. F., \& Hodgkin, S. T. 2008, \mnras,
383, 1385

\bibitem[Lodieu et al.(2007)]{lod07}
Lodieu, N., Hambly, N. C., Jameson, R. F., et al. 2007, \mnras, 374, 372




\bibitem[Luhman(2005)]{luh05usco}
Luhman, K. L. 2005, \apj, 633, L41


\bibitem[Luhman(2012)]{luh12}
Luhman, K. L. 2012, \araa, in press

\bibitem[Luhman et al.(2008)]{luh08cha}
Luhman, K. L., Allen, L. E., Allen, P. R., et al. 2008, \apj, 675, 1375

\bibitem[Luhman et al.(2007)]{luh07oph}
Luhman, K. L., Allers, K. N., Jaffe, D. T., et al. 2007, \apj, 659, 1629


\bibitem[Luhman et al.(2010)]{luh10tau}
Luhman, K. L., Allen, P. R., Espaillat, C., Hartmann, L., \& Calvet, N.
2010, \apjs, 186, 111



\bibitem[Lutz \& Lutz(1977)]{lut77}
Lutz, T. E., \& Lutz, J. H. 1977, \aj, 82, 431

\bibitem[Mart{\'\i}n(1998)]{mar98b}
Mart{\'\i}n, E. L. 1998, \aj, 115, 351

\bibitem[Mart{\'\i}n et al.(2004)]{mar04}
Mart{\'\i}n, E. L., Delfosse, X., \& Guieu, S. 2004, \aj, 127, 449 

\bibitem[Mart{\'\i}n et al.(1998)]{mar98a}
Mart{\'\i}n, E. L., Montmerle, T., Gregorio-Hetem, J., \& Casanova, S. 1998,
\mnras, 300, 733

\bibitem[Mart{\'\i}n et al.(2010)]{mar10}
Mart{\'\i}n, E. L., Phan-Bao, N., Bessell, M., et al. 2010, \aap, 517, A53

\bibitem[Merrill \& Burwell(1933)]{mer33}
Merrill, P. W., \& Burwell, C. G. 1933, \apj, 78, 87

\bibitem[Padgett et al.(2006)]{pad06}
Padgett, D. L., Cieza, L., Stapelfeldt, K. R., et al. 2006, \apj, 645, 1283

\bibitem[Palla \& Stahler(1999)]{pal99}
Palla, F., \& Stahler, S. W. 1999, \apj, 525, 772

\bibitem[Pecaut et al.(2012)]{pec12}
Pecaut M. J., Mamajek E. E., \& Bubar E. J. 2012, \apj, 746, 154

\bibitem[Prato(2007)]{pra07}
Prato, L. 2007, \apj, 657, 338

\bibitem[Prato et al.(2003)]{pra03}
Prato, L., Greene, T. P., \& Simon, M. 2003, \apj, 584, 853

\bibitem[Preibisch et al.(2002)]{pre02}
Preibisch, T., Brown, A. G. A., Bridges, T. Guenther, E., \& Zinnecker, H.
2002, \aj, 124, 404

\bibitem[Preibisch et al.(2001)]{pre01}
Preibisch, T., Guenther, E., \& Zinnecker, H. 2001, \aj, 121, 1040

\bibitem[Preibisch et al.(1998)]{pre98}
Preibisch, T., Guenther, E., Zinnecker, H., et al. 1998, \aap, 333, 619

\bibitem[Preibisch \& Mamajek(2008)]{pm08}
Preibisch, T., \& Mamajek, E. 2008,
in Handbook of Star Forming Regions, Vol. 2, The Southern Sky,
ASP Monograph Series 5, ed. B. Reipurth (San Francisco, CA: ASP), 235


\bibitem[Rebull et al.(2010)]{reb10}
Rebull, L. M., Padgett, D. L., McCabe, C.-E., et al. 2010, \apjs, 186, 259

\bibitem[Riaz et al.(2006)]{ria06}
Riaz, B., Gizis, J. E., \& Harvin, J. 2006, \aj, 132, 866

\bibitem[Riaz et al.(2009)]{ria09}
Riaz, B., Lodieu, N., Gizis, J. E. 2009, \apj, 705, 1173

\bibitem[Riaz et al.(2012)]{ria12}
Riaz, B., Lodieu, N., Goodwin, S., Stamatellos, D., \& Thompson, M. 2012,
\mnras, 420, 2497

\bibitem[Rieke et al.(2004)]{rie04}
Rieke, G. H., Young, E. T., Engelbracht, C. W., et al. 2004, \apjs, 154, 25

\bibitem[Rieke et al.(2005)]{rie05}
Rieke, G. H., Su, K. Y. L., Stansberry, J. A., et al. 2005, \apj, 620, 1010

\bibitem[Rizzuto et al.(2012)]{riz12}
Rizzuto, A. C., Ireland, M. J., \& Zucker, D. B. 2012, \mnras, 421, L97

\bibitem[Romero et al.(2012)]{rom12}
Romero, G. A., Schreiber, M. R., Cieza, L. A., et al. 2012, \apj, 749, 79

\bibitem[Scholz et al.(2007)]{sch07}
Scholz, A., Jayawardhana, R., Wood, K., et al. 2007, \apj, 660, 1517

\bibitem[Sicilia-Aguilar et al.(2006)]{sic06}
Sicilia-Aguilar, A., Hartmann, L., Calvet, N., et al. 2006, \apj, 638, 897

\bibitem[Siegler et al.(2007)]{sie07}
Siegler, N., Muzerolle, J., Young, E. T., et al. 2007, \apj, 654, 580

\bibitem[Shkolnik et al.(2009)]{shk09}
Shkolnik, E., Liu, M. C., \& Reid, I. N. 2009, \apj, 699, 649

\bibitem[Skrutskie et al.(2006)]{skr06}
Skrutskie, M., Cutri, R. M., Stiening, R., et al. 2006, \aj, 131, 1163

\bibitem[Skrutskie et al.(1990)]{skr90}
Skrutskie, M. F., Dutkevitch, D., Strom, S. E., Edwards, S., Strom, K. M., \&
Shure, M. A. 1990, \aj, 99, 1187

\bibitem[Slesnick et al.(2006)]{sle06}
Slesnick, C. L., Carpenter, J. M., \& Hillenbrand, L. A. 2006, \aj, 131, 3016

\bibitem[Slesnick et al.(2008)]{sle08}
Slesnick, C. L., Hillenbrand, L. A., \& Carpenter, J. M., 2008, \apj, 688, 377

\bibitem[Smith et al.(2008)]{smi08}
Smith, R., Wyatt, M. C., \& Dent, W. R. F. 2008, \aap, 485, 897

\bibitem[Song et al.(2012)]{son12}
Song, I., Zuckerman, B., \& Bessell, M. S. 2012, \aj, 144, 8

\bibitem[Strom et al.(1989)]{str89}
Strom, K. M., Strom, S. E., Edwards, S., Cabrit, S., \& Skrutskie, M. F. 1989,
\aj, 97, 1451

\bibitem[Sylvester \& Mannings(2000)]{syl00}
Sylvester, R. J., \& Mannings, V. 2000, \mnras, 313, 73

\bibitem[The et al.(1986)]{the86}
The, P. S., Wesselius, P. R., \& Janssen, I. M. H. H. 1986, \aaps, 66, 63

\bibitem[Torres et al.(2006)]{tor06}
Torres, C. A. O., Quast, G. R., Da Silva, L., et al. 2006, \aap, 460, 695


\bibitem[van Leeuwen(2007)]{van07}
van Leeuwen, F. 2007, \aap, 474, 653

\bibitem[Walter et al.(1994)]{wal94}
Walter, F. M., Vrba, F. J., Mathieu, R. D., Brown, A., \& Myers, P. C. 1994,
\aj, 107, 692

\bibitem[Werner et al.(2004)]{wer04}
Werner, M. W., Roellig, T. L., Low, F. J., et al. 2004, \apjs, 154, 1

\bibitem[Wright et al.(2010)]{wri10}
Wright, E. L., Eisenhardt, P. R. M., Mainzer, A. K., et al. 2010, \aj, 140, 1868

\end{thebibliography}
\end{document}